\documentclass[showpacs,showkeys,amssymb,aps,prl]{revtex4}
\newcommand{\be}{\begin{equation}} \newcommand{\ee}{\end{equation}}
\newcommand{\bea}{\begin{eqnarray}}\newcommand{\eea}{\end{eqnarray}}

\usepackage{graphicx}

\begin{document}
\hfill{SINP/TNP/2008/08}
\vspace*{10mm}
\preprint{SINP-TNP/02-23}
\title{Bound States in Graphene}
\author{Kumar S. Gupta\footnote{Email : kumars.gupta@saha.ac.in~ }}

\affiliation{ Theory Division, Saha Institute of Nuclear Physics, 1/AF Bidhannagar, Calcutta - 700064, India }

\author{Siddhartha Sen \footnote{Email : sen@maths.ucd.ie}}

\affiliation {School of Mathematical Sciences, UCD, Belfield, Dublin 4, Ireland}
\affiliation {Department of Theoretical Physics, Indian Association for the Cultivation of Science, Calcutta - 700032, India }

\begin{abstract}

We present a quantum analysis of the massless excitations in graphene with a charge impurity. When the effective charge exceeds a certain critical value,  the spectrum is quantized and is unbounded from below. The corresponding eigenstates are square-integrable at infinity and have a rapidly oscillatory behaviour in the short distance, which can be interpreted as a fall to the centre. Using a cutoff regularization, we show that the effective Coulomb interaction strength is driven to its critical value under the renormalization group flow. In the subcritical region, we find bound states with imaginary values of the energy for certain range of the system parameters. The physical significance of these bound states with imaginary eigenvalues is discussed.

\end{abstract}

\maketitle

In the tight-binding approximation, the excitations of graphene near the Fermi surface are described 
by a two dimensional (2D) massless Dirac equation \cite{semenoff, haldane}. The charge impurities in graphene follow the laws of 2D quantum electrodynamics \cite{gonz, el1,melo,novi} and vortex fields lead to fermion fractionalization \cite{hcm,jackiw}.
Recent fabrication of graphene monolayers \cite{expt} provides an opportunity to experimentally study the effects of charge impurities \cite{castro, levi1, levi2}. The massless Dirac excitations normally do not form bound states with the impurity. Numerical analysis of the coulomb impurity problem in graphene indicates the existence of bound states in the supercritical region \cite{castro}, where the Coulomb potential strength exceeds a critical value. Semiclassical analysis also suggests that an infinite family of quasibound states appear in the supercritical region \cite{levi2}. The Fermi velocity $v_F$ in graphene is approximately $10^6$ m/s and the effective fine structure constant $\alpha_g = \frac{e^2}{\hbar v_F} \sim 2.5$. This corresponds to the critical value of the effective impurity charge $\sim 1$. Measurement of local density of states using scanning tunneling spectroscopy (STS) has recently led to experimental evidence for quasibound states in this system \cite{levi2}.

In this Letter we shall present a full quantum mechanical analysis of the bound states of massless excitations in graphene in the presence of a Coulomb potential. In the supercritical regime, we shall show that there exists an infinite number of bound states. We establish that 
the spectrum is unbounded from below, which is a characteristic feature of the ''fall to the centre'' \cite{landau, case}. This feature was absent in the semiclassical analysis of \cite{levi2}. We also find that the spectrum in the supercritical region is labelled by a parameter which is not determined by the theory. The significance of this parameter can be understood as follows. The two dimensional Dirac description of graphene is a long wavelength approximation of the underlying dynamics. In the supercritical region, the power law behaviour is likely to fail  as the short distance effects become important \cite{novi}. Our result suggests that  the effect of the breakdown of the power law potential at short distance can be parametrized by a single quantity. This parameter is not predicted by the theory itself, but must be determined empirically. The ratio of the eigenvalues is however independent of this parameter.

A special feature of graphene with charge impurities is that the nonlinear screening effects drives the effective charge to the critical value \cite{levi1}. In our formalism this is also present and can be interpreted as a renormalization group flow, which is well known for singular inverse square interactions \cite{rajeev}. We will show that the corresponding beta function has an ultraviolet stable fixed point at the critical value of the Coulomb potential. 

For the massless two dimensional Dirac equation, there is no bound state in the subcritical region. This statement is true only for real value of the energy. We find that for certain range of the parameters, the system admits square-integrable bound states with eigenvalues $\pm i$. Using von Neumann's theory of self-adjoint extensions \cite{reed}, we show that these states lead to a one parameter family of boundary conditions for the relevant range of system parameters. These generalized boundary conditions are expected to affect the phase shifts and the S-matrix in the scattering sector. This situation is analogous to the occurrence of anomalies for singular potentials \cite{ano1,ano2}. 

The 2D massless Dirac equation in the presence of a Coulomb potential $V(\rho) = {Z e^2 \over \rho}$ can be written as
\be \label{dirac}
\hbar v_F 
\left( \begin{array}{cc}
0 & - i \partial_x - \partial_y  \\
- i \partial_x + \partial_y & 0 \\
\end{array} \right) \chi 
= [E - V(\rho)] \chi,
\ee
where $\rho$ is the radial coordinate on the two dimensional x-y plane. Let $\phi$ denote the corresponding polar angle on the plane. 
Following \cite{levi1}, the separation of variables can be performed using the ansatz 
\be \label{separ} 
\chi (\rho, \phi) = \left( \begin{array}{c}
{F (\rho ) ~ \Phi_m(\phi) } \\
{ G (\rho )~ \Phi_{m+1}(\phi) } \\
\end{array} \right)
\equiv
\left( \begin{array}{c}
{[u (\rho ) + v (\rho)] ~ \Phi_m(\phi) } \\
{ [u (\rho) - v (\rho)]~ \Phi_{m+1}(\phi) } \\
\end{array} \right) \rho^{s- \frac{1}{2}}  e^{i k \rho},
\ee
where $ \Phi_m(\phi) = \frac{1}{\sqrt{2 \pi}} e^{i(m - \frac{1}{2}) \phi}$, $m$ is the half integer azimuthal quantum number and 
\be \label{para}
s = \sqrt{m^2 - \beta^2}, ~~~k = - \frac{E}{\hbar v_F}, ~~~~ \beta = \frac{Z e^2}{\kappa \hbar v_F},
\ee
with $\kappa$ as the effective dielectric constant.  The minimum magnitude of the azimuthal quantum number is $m = \frac{1}{2}$ and 
from (\ref{para}) we see that the  corresponding critical value of $\beta$ is given by $\beta_c = \frac{1}{2}$.  Assuming that graphene electrons are the only source of screening gives $\kappa \approx 5$ \cite{gonz}, which together with $\beta_c = \frac{1}{2}$, gives the critical value of $Z$ as $Z_c \sim 1$ \cite{levi1}. This feature makes the system interesting from an experimental point of view.

The Dirac equation (\ref{dirac}) with the ansatz (\ref{separ}) leads to two coupled equations,
\bea
\rho \frac{d u}{d \rho} + (s + i \beta + 2ik \rho)u - mv &=& 0, \\
\rho \frac{d v}{d \rho} + (s - i \beta) v - m u &=& 0 .
\eea
Defining $z = -2ik\rho$, from (4) and (5) we get
\be \label{veqn}
z \frac{d^2 v}{d z^2} + (1 + 2s -z) \frac{d v}{d z} - (s - i \beta) v = 0,
\ee
which is of the form of Kummer's equation for the confluent hypergeometric function \cite{as}.  
We now define a new function $\eta$ as
\be \label{xform}
v = \rho^{-(\frac{1}{2} + s)}~e^{-i k \rho}~ \eta.
\ee
Using (\ref{xform}) in (\ref{veqn}), we get
\be \label{whit}
\frac{d^2 \eta}{d z^2} + \left [ - \frac{1}{4} + \frac{\alpha}{z} + \frac{(\frac{1}{4} - s^2)}{z^2} \right ] \eta = 0,
\ee
where $\alpha = \frac{1}{2} + i \beta $. Eqn. (\ref{whit}) has the form of Whittaker's equation \cite{as}. 

We shall first analyze the supercritical case where $\beta > \beta_c$.  In this case, $s = \pm i \mu$, $\mu = \sqrt{\beta^2 - m^2}$ and Eqn. (\ref{whit}) can be written as 
\be \label{whit1}
\frac{d^2 \eta}{d z^2} + \left [ - \frac{1}{4} + \frac{\alpha}{z} + \frac{(\frac{1}{4} + \mu^2)}{z^2} \right ] \eta = 0.
\ee
The general solution of (\ref{whit1}) is given by
\be \label{soln}
\eta (z)  = A_1 e^{i \theta} M_{\alpha, i \mu} (z) + A_2 e^{-i \theta} M_{\alpha, -i \mu} (z),
\ee
where $A_1$ and $A_2$ are constants and 
\be \label{soln1}
M_{\alpha, \pm i \mu} = e^{-\frac{z}{2}} z^{\frac{1}{2}  \pm i \mu} M (\pm i \mu - i \beta, 1 \pm 2 i \mu, z),
\ee
with $M$ on the right hand side of (\ref{soln1}) denoting the confluent hypergeometric function \cite{as}. 

We shall now construct solutions which are square integrable at infinity. Let us first note that for the functions $F (\rho)$ and $G(\rho)$ in 
(\ref{separ}), the measure of integration is given by $\rho d \rho$. Now consider the contribution of the function $\eta$ in (\ref{soln}) to the quantity $ \int |F|^2 \rho d \rho $ at asymptotic infinity. Using the asymptotic formula (13.5.1 of \cite{as}) for the confluent hypergeometric function, we find that as $|z| = |-2ik \rho | \rightarrow \infty$, 
\be \label{asym1}
M_{\alpha,  \pm i \mu} (z) \longrightarrow e^{ik \rho} e^{ \pi ( \pm \mu - \beta)} (-2ik\rho)^{\frac{1}{2} + 
i \beta} \frac{\Gamma(1 \pm 2 i \mu)}{\Gamma(1 \pm i \mu + i \beta)}
+  e^{-ik \rho} (-2ik\rho)^{-\frac{1}{2} - i \beta} \frac{\Gamma(1 \pm 2 i \mu)}{\Gamma( \pm i \mu - i \beta)}.
\ee
Using (\ref{separ}), (\ref{xform}) and (\ref{asym1}), it is easy to see that as $\rho \rightarrow \infty $, the contribution of the first term on the rhs of (\ref{asym1}) to $|F|^2$ behaves as $\sim \frac{1}{\rho}$ while that from the second term behaves as $\sim \frac{1}{\rho^3}$. Hence, the first term in the rhs of (\ref{asym1}) provides a divergent contribution to $ \int |F|^2 \rho d \rho $ while that of the second term is convergent. We now consider the part of the solution $\eta$ in (\ref{soln}), arising from the first term in the rhs of (\ref{asym1}), which leads to the divergence of the norm of $F$.
This is denoted by $\eta_{{\rm{div}}}$ and is given by
\be
\eta_{{\rm{div}}} \rightarrow e^{i k \rho} (-2ik\rho)^{\frac{1}{2} + i \beta} e^{ - \pi \beta }
\left [ A_1 e^{i \theta} e^{ \pi \mu } \frac{\Gamma(1 + 2 i \mu)}{\Gamma(1+ i \mu + i \beta)} +
A_2 e^{-i \theta} e^{ - \pi \mu } \frac{\Gamma(1 - 2 i \mu)}{\Gamma(1 -i \mu + i \beta)} \right ].
\ee
We now choose $A_2 = A_1 e^{ 2 \pi \mu } \frac{\Gamma(1 -i \mu + i \beta)}{\Gamma(1 - i \mu - i \beta)}$ and denote 
$ \frac{\Gamma(1 + 2 i \mu)}{\Gamma(1 + i \mu + i \beta)} 
= \xi e^{i \gamma}$. Therefore, as $ \rho \rightarrow \infty$, we get
\be
\eta_{{\rm{div}}} \rightarrow \rightarrow 2 A_1 e^{i k \rho} (-2ik\rho)^{\frac{1}{2} + i \beta} e^{  \pi (\mu - \beta) }
\xi \cos (\theta + \gamma ).
\ee
We can make $\eta_{{\rm{div}}} = 0$ if we impose the condition that 
\be \label{quant1}
\cos (\theta + \gamma ) = 0 ~~~ {\rm{or}} ~~~ \theta = -\gamma + \left ( n + \frac{1}{2} \right ) \pi,~~ n \in Z.
\ee
This gives a quantization of the parameter $\theta$. The quantization of $\theta$ ensures that as 
$ \rho \rightarrow \infty$, $ \int |F|^2 \rho d \rho $ will remain finite. Knowing $\eta$, the components $u$ and $v$ of the solution to the Dirac equation can be obtained from (\ref{xform}) and (5). They are square-integrable and represent the bound state solution. 

We now proceed to obtain the quantized energy levels. For that purpose, consider the short distance limit of (\ref{soln}). Denoting 
$ \frac{\Gamma(1 -i \mu + i \beta)}{\Gamma(1 - i \mu - i \beta)} = C e^{i 2 \delta}$, we get, as $z \rightarrow 0$, 
\be \label{short}
\eta \sim  \sqrt{z} \left [ (1 + C) \cos (\theta - \delta + \mu \ln z) + (1-C) \sin (\theta - \delta + \mu \ln z) \right ] .
\ee
The short distance behaviour of the system can also be inferred by looking at the corresponding indical equation. In this case, the roots of the indical equation are given by $ \frac{1}{2} \pm i \mu $. Using these two roots, the wavefunction at short distance can be obtained as
\bea \label{indic}
\eta &\sim& \sqrt{-2i\rho} \left [ e^{i B} e^{i \mu \ln (-2i\rho)} + C e^{-i B} e^{-i \mu \ln (-2i\rho)} \right ] \nonumber \\
&\sim& \sqrt{-2i\rho} \left [ (1+C) \cos (B + \mu \ln (-2i\rho)) + i (1-C) \sin (B + \mu \ln (-2i\rho)) \right ],
\eea
where $B$ is a real constant. The short distance behaviour of the wavefunction obtained from (\ref{short}) and (\ref{indic}) must match. This is possible provided
\be \label{q1}
\theta + \delta + \mu \ln k = B. 
\ee
Thus we obtain the quantized energy eigenvalues as 
\be \label{q2}
E_n = - \frac{k_n}{\hbar v_F} =  - \frac{e^{- (n + \frac{1}{2}) \frac{\pi}{\mu} + A }}{\hbar v_F}, ~~n \in Z,
\ee 
where $A = \frac{(B-\delta)}{\mu}$ is a constant. The quantity $\delta$ depends on the system parameters $\mu$ and $\beta$. $B$ on the other hand is a real constant which is not fixed by the theoretical analysis. This leads to a one parameter family of inequivalent spectra in the supercritical regime. As mentioned before, $B$  encodes the effect of short distance physics, which are expected to be important as we approach length scales of the order of the lattice spacings. The analysis suggests that it is not important to know the the details of the short distance interactions and their effect on the spectrum appears through a single parameter $B$, which should be determined empirically. The ratio of the various energy levels are independent of the parameter $B$. It may be noted that the above quantization has been carried out with only the function $v(\rho)$. The function $u(\rho)$ can be determined from (5) and it can be shown that it does not change the quantization condition.

We have thus found an infinite number of bound states in graphene containing a impurity, when the effective charge exceeds the critical value. These states are square-integrable at infinity and have a rapidly oscillatory behaviour in the short distance limit, which is a feature of the fall to the centre. Our analysis is fully quantum and nonperturbative, valid for any value of $\beta > \beta_c$. In that sense we go beyond the treatment in \cite{levi2}. The semi-classical analysis in \cite{levi2} yields quantized energy levels only for $n > 0$ whereas we obtain quantized energy levels for all integer values of $n$. Due to this, the spectrum we get is unbounded from below, which is an important feature of the supercritical region. We also find an accumulation point at zero energy, which is seen in the exact numerical analysis of the system as well \cite{castro}.
 
Another interesting feature of graphene is that the screening in the supercritical region drives the effective charge to the critical value \cite{levi1}. In our formulation this is achieved when $\mu \rightarrow 0$ from the supercritical region. Within the scope of the quantum mechanics discussed here, this effect can be viewed as a renormalization group flow arising from the short distance effects. In the short distance limit, the eigenvalue equation (\ref{whit1}) assumes the conformal form given by
\be \label{conform}
\frac{d^2 \eta}{d \rho^2} +  \frac{(\frac{1}{4} + \mu^2)}{\rho^2} \eta = \lambda \eta,
\ee
with the eigenvalue $\lambda=0$.  Eqn. (\ref{conform}) for a general $\lambda$ has been analyzed in \cite{rajeev} and provides an example of renormalization in quantum mechanics. In the supercritical region, the spectrum is unbounded from below. In order to regulate this divergence, we introduce a short distance cutoff at $\rho=a$ and impose the boundary condition that the wave-function vanishes below the cutoff. With such a boundary condition, the eigenvalues of (\ref{conform}) can be calculated analytically for small values of $\mu$, leading to a finite bound state spectrum for $\lambda$ given by \cite{rajeev}
\be \label{en}
\lambda_n = -e^{-\frac{2n\pi}{\mu}} \left [ \frac{2}{ae^\sigma} \right ] ,
\ee
where $\sigma$ is the Euler's constant and $n=1,2,....,\infty $. Note that for any finite value of the cutoff, as $n \rightarrow \infty$, $\lambda_n \rightarrow 0$ and the zero eigenvalue depicts an accumulation point for this system. In this limit we recover the short distance form of (\ref{whit1}). The eigenvalues $\lambda_n$ also explicitly depend on the cutoff and they diverge as the cutoff is removed. This indicates that the breakdown of the 2D Dirac description of graphene at short distances \cite{novi}. In the spirit of renormalization group analysis, we now make the coupling $\mu$ a function of the cutoff, i.e. take $\mu = \mu (a)$. In order to find the dependence of $\mu$ on the cutoff, we demand that the zero eigenvalue remains unchanged as the cutoff is removed. This leads to a beta function for the coupling $\mu$ given by \cite{rajeev}
\be \label{bet1}
{\tilde {\beta}} (\mu) = -a \frac{d \mu}{d a} \approx - \mu^2.
\ee
This beta function has an ultraviolet stable fixed point at $\mu=0$. In other words, the effective charge is driven to its critical value as the cutoff is removed. We therefore recover the same result obtained from the analysis of nonlinear effects on screening \cite{levi1}.  

We shall now analyze graphene with charge impurity in the subcritical region with $\beta < \beta_c$. $s$ is now a positive real number, $s=0$ denoting the critical value.  In what follows we shall restrict our attention to $s > 0$. If the corresponding Dirac equation (\ref{dirac}) had a mass term, there would be bound states in the subcritical region, with the bound state energy proportional to the mass \cite{novi}. In our case we do not have the mass term in (\ref{dirac}) and would normally not expect any bound state, at least with real energy. 
For the moment, let us consider the system where the quantity $k$ is not real. In particular, we ask the question whether there exist bound states with values of $k = \pm i$. We shall discuss the significance of such solutions later.  

Consider first the case with $k = + i$. In this case, $z= -2ik\rho = 2 \rho$ and a possible solution of (\ref{veqn}) is given by 
\be \label{vplus}
v_+ = U (s-i \beta, 1 + 2s, z) = U (s-i \beta, 1 + 2s, 2\rho).
\ee
From (5), we get that
\be \label{uplus}
u_+ = \frac{(s-i\beta)}{m} \left [ U (s-i \beta, 1 + 2s, 2\rho) - 2 \rho  U (s-i \beta + 1, 2 + 2s, 2\rho ) \right ],
\ee
where we have used the relation $U^{\prime}(a,b,z) = -a U(a+1, b+1,z)$ \cite{as}. Using (\ref{vplus}) and (\ref{uplus}) in 
(\ref{separ}), we obtain
\be \label{Fplus}
F_+ (\rho) = e^{-\rho}  \left [ \frac{(m+ s-i\beta)}{m} \rho^{(s - \frac{1}{2})} U (s-i \beta, 1 + 2s, 2\rho) 
-   \frac{2(s-i\beta)}{m} \rho^{(s + \frac{1}{2})} U (s-i \beta + 1, 2 + 2s, 2\rho )  \right ] 
\ee
Since $\rho$ is positive, as $\rho \rightarrow \infty$, $U(a,b,2\rho) \sim \rho^{-a}$. Therefore we see that as $\rho \rightarrow \infty$, 
$F_+(\rho) \rightarrow 0$. Next consider the short distance limit of $F_+(\rho)$. Using the relation 
\be \label{U}
U(a,b,y) = \frac{\pi}{\sin(\pi b)} \left [ \frac{M(a,b,y)}{\Gamma(1+a-b)\Gamma(b)} -y^{1-b} \frac{M(1+a-b,2-b,y)}{\Gamma(a)\Gamma(2-b)} \right ]
\ee
and noting that as $y \rightarrow 0$, $M(a,b,y) \rightarrow 1$ \cite{as}, we get that as $\rho \rightarrow 0$,
\be \label{Fplush}
\int |F_+|^2 \rho d \rho  \sim \int \rho^{-2s} d \rho + ~{\mathrm {other ~ convergent ~ terms}}.
\ee
Thus, for the range $0 < s < \frac{1}{2}$, we see that $F_+ (\rho)$ is a square integrable function. Similar analysis shows that for $0 < s < \frac{1}{2}$, the entire radial wave-function is square integrable. Thus, for $k=i$, we have a single square integrable bound state when 
$0 < s < \frac{1}{2}$. We denote the corresponding radial bound state with $\psi_+$.

Let us now consider the case when $k=-i$, where $z=-2ik\rho = - 2 \rho$. In this case, a possible solution of (\ref{veqn}) is given by \cite{as}
\be \label{vmin}
v_- = e^z U(1+s+i\beta, 1+2s, -z) = e^{-2\rho} U(1+s+i\beta, 1+2s, 2\rho).
\ee
From (5), we get that
\be \label{umin}
u_- = e^{-2\rho} \left [ \frac{(s-i\beta)}{m} U(1 + s + i\beta, 1+2s, 2 \rho) -2 \rho \frac{(2+s-i\beta)}{m} U(2+s+i\beta, 2+2s, 2\rho) \right ].
\ee
From these we see that
\be \label{Fmin}
F_- (\rho) = e^{-\rho}  \left [ \frac{(m+ s-i\beta)}{m} \rho^{(s - \frac{1}{2})} U (1+s+i \beta, 1 + 2s, 2\rho) 
-   \frac{2(2+s-i\beta)}{m} \rho^{(s + \frac{1}{2})} U (s-i \beta + 2, 2 + 2s, 2\rho )  \right ] 
\ee
From (\ref{Fmin}) we notice that as $\rho \rightarrow \infty $, $F_- (\rho) \rightarrow 0$. In the short distance limit, we can again show that $F_-$ and the corresponding entire radial wave-function is square integrable when $0 < s < \frac{1}{2}$. Thus, for $k=-i$ also, we have a single square integrable bound state when $0 < s < \frac{1}{2}$. We denote the corresponding radial bound state with $\psi_-$.

We now comment on the physical significance of the square-integrable bound states $\psi_{\pm}$ with $k= \pm i$, which exist when $0 < s < \frac{1}{2}$. 
Their significance can be understood following von Neumann's treatment of self-adjoint operators in quantum mechanics \cite{reed}. The original radial Dirac operator for graphene, denoted by $H_\rho$ is defined with the boundary condition that the wave-function vanishes at the location of the charge impurity given by $\rho=0$. This boundary condition defines a domain of the Dirac operator denoted by $D (H_\rho)$, whose elements vanish at $\rho = 0$. The corresponding adjoint operator $H^{\dagger}_\rho$ in this case has the same differential expression as $H_\rho$. Let $n_{\pm}$ denote the number of linearly independent square integrable solutions of the eigenvalue equation for $H^{\dagger}_\rho$, with eigenvalues $\pm i$. In our case, we have found that $n_+ = n_- = 1$ when $0 < s < \frac{1}{2}$ and $n_{\pm} = 0$ otherwise. The theory due to von Neumann \cite{reed} implies that for $0 < s < \frac{1}{2}$, the radial Dirac operator $H_\rho$ is not self-adjoint in the domain $D (H_\rho)$. It furthermore says that $H_\rho$ would be self-adjoint in a different domain, denoted by $D_\nu (H_\rho)$, whose elements can be written as $\psi + Q (\psi_+ + e^{i \nu} \psi_-)$ where $\psi \in D (H_\rho)$, $Q$ is a constant and $\nu \in R$ (mod $2\pi$). In effect, this provides a one parameter family of self-adjoint extensions, or equivalently boundary conditions, which are labelled by a single real parameter $\nu \in [0,2\pi]$. Thus, the physical significance of the bound states with $k = \pm i$ is that they lead to a more general class of boundary conditions which are allowed by the rules of quantum mechanics. The quantity $\nu$ which classifies the different boundary conditions is called the self-adjoint extension parameter. It is expected that the phase shifts and the S-matrix in the scattering sector of the Dirac operator (\ref{dirac}) would depend explicitly on the parameter $\nu$ for $0 < s < \frac{1}{2}$ \cite{ano2}. Thus, even though the solutions with $k = \pm i$ are not physical bound states, they do play a role in determining the spectrum of the problem when $0 < s < \frac{1}{2}$.

In summary, we have presented a quantum analysis of bound states in graphene in the presence of a charge impurity. In the supercritical region, we find a quantized spectrum with the energy unbounded from below. The eigenstates are square-integrable at infinity and have rapidly oscillatory behaviour at short distances, indicating a fall to the centre. They are parametrized by a single real constant, which captures the effect of short distance physics. Experimental evidence for the bound states in the supercritical regime was found in \cite{levi2}. It is expected that the states described here correspond to what was observed experimentally. We also show the Coulomb potential is driven to its critical value under a renormalization group flow. Finally, for certain range of the system parameters in the subcritical region, we find bound states with eigenvalues given by $\pm i$. These states determine the allowed boundary conditions for this system and the corresponding self-adjoint extension parameter is expected to affect the phase shifts and the S-matrix.

KSG would like to thank the School of Theoretical Physics at the Dublin Institute for Advanced Studies for hospitality, where a part of this work was done. We thank V. M. Pereira and A. H. Castro Neto for bringing ref. \cite{castro} to our attention.


\begin{thebibliography}{99}
 
\bibitem{semenoff} G. W. Semenoff, Phys. Rev. Lett. {\bf 53}, 2449 (1984).

\bibitem{haldane} F. D. M. Haldane, Phys. Rev. Lett. {\bf 61}, 2015 (1988).

\bibitem{gonz} J. Gonzalez, F. Guinea and M. A. H. Vozmediano, Nucl. Phys. {\bf B 424}, 595 (1994); J. Low Temp. Phys. 
{\bf 99}, 287 (1995).

\bibitem{el1} M. I. Katnelson, K. S. Novoselov and A. K. Geim, Nature Phys. {\bf 2}, 620 (2006).

\bibitem{melo} D. P. DiVincenzo and E. J. Melo, Phys. Rev. {\bf B 29}, 1685 (1984).

\bibitem{novi} D. S. Novikov, Phys. Rev. {\bf B 76}, 233402 (2007).

\bibitem{hcm} C-Y. Hou, C. Chamon and C. Mudry, Phys. Rev. Lett. {\bf 98}, 186809 (2007).

\bibitem{jackiw} R. Jackiw and S.-Y.Pi, Phys. Rev. Lett. {\bf 98}, 266402 (2007).

\bibitem{expt} K. S. Novoselov, A. K. Geim, S. V. Morozov, D. Jiang, Y. Zhang, S. V. Dubonos, 
I. V. Grigorieva and A. A. Firsov, Science {\bf 306}, 666 (2004).

\bibitem{castro} V. M. Pereira, J. Nilsson and A. H. Castro Neto, Phys. Rev. Lett. {\bf 99}, 166802 (2007).

\bibitem{levi1} A. V. Shytov, M. I. Katsnelson and L. S. Levitov, Phys. Rev. Lett. {\bf 99}, 236801 (2007).

\bibitem{levi2} A. V. Shytov, M. I. Katsnelson and L. S. Levitov, Phys. Rev. Lett. {\bf 99}, 236802 (2007).

\bibitem{landau} L. D. Landau and E. M. Lifshitz, {\it Quantum Mechanics} (Pergamon, London, 1958.)

\bibitem{case} K. M. Case, Phys. Rev. {\bf 80}, 797 (1950).

\bibitem{rajeev} K. S. Gupta and S. G. Rajeev, Phys. Rev. {\bf D48}, 5940 (1993).

\bibitem{reed} M. Reed and B. Simon, {\it Methods of Modern Mathematical
Physics}, volume 2, (Academic Press, New York, 1972).

\bibitem{ano1} J. G. Esteve, Phys. Rev. {\bf D 34}, 674 (1986); Phys. Rev. {\bf D 66}, 125013 (2002).

\bibitem{ano2} B. Basu-Mallick, P. K. Ghosh and Kumar S. Gupta, Nucl. Phys. {\bf B 659}, 437 (2003); Phys. Lett. {\bf A 311}, 87 (2003). 

\bibitem{as} M. Abramowitz and I. A. Stegun, {\it Handbook of Mathematical Functions} (Dover, New York, 1964.)

\end{thebibliography}
\end{document}